\documentstyle[prb,aps,epsfig]{revtex}

\begin{document}

\draft


\title{Low-Temperature Magnetic Penetration Depth in $d$-Wave Superconductors:
Zero-Energy Bound State and Impurity Effects}

\author{Yu.~S.~Barash{$^{1}$}, M.~S.~Kalenkov{$^{1}$}, and
J.~Kurkij\"arvi{$^{2}$}}
\address{$^{1}$ P.N. Lebedev Physical Institute, Leninsky Prospect 53,
Moscow 117924, Russia\\
$^{2}$ Department of Physics, \AA bo Akademi, Porthansgatan 3, FIN-20500
\AA bo, Finland}




\maketitle


\begin{abstract}
We report a theoretical study on the deviations of the Meissner penetration
depth $\lambda(T)$ from its London value in $d$-wave superconductors at low
temperatures. The difference arises from low-energy surface Andreev bound
states. The temperature dependent penetration depth is shown to go through a
minimum at the temperature $T_{m0}\sim \sqrt{\xi_0/\lambda_0}T_c$ if the
broadening of the bound states is small. The minimum will straighten out when
the broadening reaches $T_{m0}$. The impurity scattering sets up the
low-temperature anomalies of the penetration depth and destroys them when the
mean free path is not sufficiently large. A phase transition to a state with a
spontaneous surface supercurrent is investigated and its critical temperature
determined in the absence of a subdominant channel activated at low
temperatures near the surface. Nonlinear corrections from Andreev low-energy
bound states to the penetration length are obtained and shown, on account of
their broadening, to be small in the Meissner state of strong type II
superconductors.
\end{abstract}





\section{Introduction}
The low-temperature behavior of the magnetic penetration length in $d$-wave
superconductors is in general a great deal more complicated than that of their
isotropic $s$-wave cousins. The changing sign of the order parameter, according
to where one looks on the Fermi surface, entails coherent zero-energy or
low-energy bound states in $d$-wave superconductors localized at smooth or
almost smooth surfaces or
interfaces\cite{buch81,hu94,tan95,buch95,fog97,bbs97}.  These bound states
feature peculiar low-temperature contributions to the magnetic penetration
length\cite{wal98}(see also\cite{alf981}) and the zero-bias conductance
peak\cite{hu94,tan95,fog97}(see also
\cite{cov97,alf97,ek97,ap98,alf981,alf982,sin98,wei981,wei982,apr99,deutch99}).

A minimum in the penetration depth of $YBa_2Cu_3O_{7-\delta}$ films\cite{wal98}
and grain boundary junctions\cite{alf981} was thus interpreted as evidence for
low-energy Andreev bound states. A conventional shielding-current contribution
to the Meissner effect would obviously just monotonically reduce the
penetration depth when the temperature goes down. On the other hand, a
paramagnetic contribution from low-energy bound states increases the
penetration depth. The interplay of these two effects amounts to a minimum in
the penetration depth as a function of the temperature. The characteristic
temperature $T_{m0}$ of this anomaly is shown to be the order
$\sqrt{(\xi_0/\lambda_0)}T_c\ll T_c$ if the broadening $\gamma$ of the bound
states is sufficiently small. At this temperature region the bound state
contribution to the penetration depth competes with the low-temperature
correction from shielding supercurrents to its zero-temperature value.

An alternative explanation of an upturn in the penetration depth is possible in
compounds whose bulk paramagnetic properties grow when the temperature goes
down, like in the electron-doped cuprate superconductor
$Nd_{1.85}Ce_{0.15}CuO_{4-y}$. There the paramagnetism arises from $Nd^{3+}$
ions\cite{coop96,alf99,proz00,tsuei00}. We will not discuss these compounds
below.

There is yet another important temperature associated with the magnetic
penetration depth, $T_{s}\sim(\xi_0/\lambda_0)T_c$. If a given crystal
orientation does not carry quasiparticle Andreev bound states, a nonlocal
effect can take over as a correction to the zero temperature penetration depth
in the clean limit\cite{kos97}. Then in other orientations which do admit
Andreev states, the bound-state contribution and the spontaneous surface
supercurrent in particular, can in turn overwhelm the nonlocal effect. At
$T\sim T_{s}$ the bound state paramagnetic contribution to $\lambda$ in the
clean limit\cite{clean} is the order of the total London penetration depth
$\lambda_0$ from the screening currents. In the absence of sub-dominant pairing
channels, a spontaneous surface supercurrent brought about by the bound states
may arise below the temperature $T_{s}$\cite{hig97,sigr99} (see
also\cite{belz99} on a similar effect of spontaneous magnetization brought
about by low energy interface bound states). Having in mind high-temperature
superconducting compounds, we will discuss strong type II superconductors.
Then $(\xi_0/\lambda_0)$ is easily the order $0.01$, and the low temperature
range splits up into at least three areas staked out by $0$, $T_{s}$ and
$T_{m0}$ ($T_{s}\sim (\xi_0/\lambda_0)T_c\ll T_{m0}\sim\sqrt{(\xi_0/\lambda_0)}
T_c\ll T_c$). Quasiparticle scattering off impurities or surface roughness and
inelastic processes may also play an important role if they bring about a
broadening $\gamma$ of the bound states the order or greater than the
characteristic temperature $T_{m0}$ ($T_s$).

We assume below that nonmagnetic impurities dominate the scattering and the
broadening. Nonmagnetic impurities in superconductors with an anisotropic order
parameter are known to be pair breaking. They suppress $T_c$ analogously to
what happens to isotropic superconductors with magnetic impurities. Assuming
superconductors always clean within the conventional definition $\xi\ll l$, we
disregard this kind of effects throughout the article. Even then impurity
broadening of Andreev bound states in anisotropically paired superconductors
can be significant. Since the broadening removes singularities in the density
of states (for instance, $\delta$-peaks from quasiparticle bound states) as
well as in other related physical quantities, superconductors can be sensitive
to extremely small concentrations of impurities\cite{clean}. This is analogous
to the role of pair breaking and small anisotropy of the gap in the Riedel
anomaly in isotropic $s$-wave superconductors. The Riedel anomaly is associated
with the BCS singularity in the density of states. Pair breaking and small
anisotropy of the gap are known to wipe out the BCS singularity in the density
of states averaged over the Fermi surface, and control the height of the Riedel
peak\cite{barone82}.

The emphasis of the present work is on the various effects of broadening on the
low temperature anomalies of the Meissner effect. The zero-energy pole-like
term of what is known as the quasiclassical Green's function was exploited in
the investigation. Broadening is introduced into the pole-like term simply
sliding the pole along the imaginary energy axis. With small broadening,
relatively simple expressions are found for the penetration length in the two
lowest-temperature regions defined above. If $T_{s(m0)}\lesssim\gamma\ll T_c$,
the growing $\gamma$ can wipe out the low-temperature anomalies. Beginning with
the critical broadening $\gamma_{s(m)}$, anomalies at $T_{s(m)}$ are fully
destroyed. It turns out that unitary scatterers need to come with
significantly larger scattering rates $\Gamma_{s(m)}$ than Born impurities in
order to achieve the critical broadening $\gamma_{s(m)}$. This effect is
peculiar of the impact of impurities on the Andreev bound states as seen in the
local density of states and Josephson critical currents\cite{pbbi99}. For this
reason, the requirements the mean free path must meet for the low temperature
anomalies to show up are sensitive to the strength of the impurity potential
and very different in the unitary and the Born limits.

For Born scatterers, the shortest normal-state impurity mean free path $l$
which preserves the low temperature upturn at $T\sim T_m$ is shown to be
$\lambda_0\lesssim l$. This looks quite restrictive although conceivably
compatible with the strikingly large low temperature mean free paths in some
high-$T_c$
compounds\cite{bonn92,bonn94,krish95,krish99,hosseini99,berlinsky00}. For the
spontaneous surface supercurrent in the absence of a subdominant component at
the surface, we find the threshold $\lambda^2_0/\xi_0\lesssim l$. This demands
extraordinary clean samples not available for the time being. On the other
hand, the requirements set by unitary scatterers are much weaker and probably
can be met. In this case surface roughness is likely to control the broadening
and the experimental observability of the effects.

We also examine what the Andreev bound states do to the nonlinear Meissner
effect. At low temperatures $T\ll T_c$, the field $\widetilde{H}_0$ at which
the nonlinear response of the bound states saturates in the clean
limit\cite{clean} is much weaker than the one from the screening current.
Ignoring the broadening, $\widetilde{H}_0$ is a linear function of the
temperature. With $T\lesssim T_s$, nonlinear effects from the bound states
become important already in the Meissner state. Close to the transition to the
state with a spontaneous surface supercurrent, a nonlinear term entering into
the Landau mean-field free energy is important also in a weak external field.
The broadening $\gamma$ introduces another field, $\widetilde{H}_{\gamma}$
characterizing the nonlinear consequences of the bound states at $\pi
T\lesssim\gamma$. For sufficient broadening $\pi T_s\ll \gamma$, we get
$H_{c1}\ll\widetilde{H}_{\gamma}$ and the nonlinear terms are shown always to
be small in the Meissner state.

\section{The upturn in the low-temperature dependence of the penetration depth}

Our considerations are based on the quasiclassical matrix Green function which
describes quasiparticle excitations in thermal equilibrium. The quasiclassical
propagator  $\hat{g}(\bbox{p}_f,x,\varepsilon_n)$ satisfies Eilenberger's
equations, which have a $2\times 2$ particle-hole matrix form
$$
\left[\left(i\varepsilon_n+\frac{\displaystyle e}{\displaystyle c}
{\rm{\bf v}}_f\hspace*{-2pt} \cdot \hspace*{-2pt}{\bf A}({\bf R})\right)
\hat{\tau}_3-\hat{\Delta}(\bbox{p}_f,{\bf R})-\hat{\sigma}(\bbox{p}_f,{\bf R};
\varepsilon_n),
\hat{g}(\bbox{p}_f,{\bf R};\varepsilon_n)\right]+
$$
\begin{eqnarray}
+i {\rm{\bf v}}_f
\hspace*{-2pt} \cdot \hspace*{-2pt}
\bbox{\nabla} \hspace*{-2pt}_{{\bf R}}\:
\hat{g}(\bbox{p}_f,{\bf R};\varepsilon_n)=0 \enspace , \label{eil} \\
\hat{g}^2(\bbox{p}_f,{\bf R};\varepsilon_n)=-\pi^2\hat{1} \enspace ,
\qquad\qquad\qquad
\label{norm}
\end{eqnarray}
where $\varepsilon_{n}=(2n+1)\pi T$ are the Matsubara energies, $\bbox{p}_f$
the momentum on the Fermi surface, ${\rm{\bf v}}_f$ the Fermi velocity,
${\bf A}$ the vector potential, $\hat\Delta$ the order parameter matrix,
and $\hat{\sigma}$ the impurity self-energy. A symbol with a `hat' denotes a
matrix in the Nambu space.

The propagator $\hat g$ and the order parameter matrix $\hat\Delta$
parameterize as
\begin{eqnarray}
\hat g=\left(\begin{array}{cr}
                                   g   &  f \\
                   f^+ & -g
        \end{array}\right)\enspace
\qquad \mbox{and} \qquad
\hat \Delta=\left(\begin{array}{cc}
                                       0     &  \Delta \\
                   -\Delta^* &     0
        \end{array}\right)\enspace.
\label{matrix}
\end{eqnarray}
The gap function $\Delta(\bbox{p}_f,{\bf R})$ is related to the
anomalous Green function $f$ and must be determined self-consistently.
The diagonal part $g(\bbox{p}_f,{\bf R},\varepsilon_n)$ of the
full matrix propagator $\hat g$ carries information on the
electrical current density
\begin{equation}
\bbox{j}({\bf R})=2eTN_f\sum\limits_{\varepsilon_n}\Big<{\rm\bbox{v}}_fg(
\bbox{p}_f,{\bf R},\varepsilon_n)\Big>_{S_f}
\enspace .
\label{gcur}
\end{equation}
Here $N_f$ is the normal state density of states per spin direction and
\mbox{$<\ldots >_{S_f}$} means averaging over quasiparticle states at the Fermi
surface.

Let an anisotropic singlet strong type II superconductor occupy the right
half-space $x>0$. A magnetic field is applied along the $z$-axis. The induced
supercurrent and the vector potential (in the gauge ${\rm div}{\bf A}(
{\bf R})=0$ and vanishing in the bulk) have only $y$-components.
The linear current-field relation in general has a nonlocal form, i.e. $j(x)
=-\int\nolimits_0^{+\infty} Q(x,x',T) A(x') dx'$.

For strongly type II superconductors with nodes in the order parameter,
a nonlocal current-field relation can be of importance only at very low
temperatures $T\lesssim T_{s}$\cite{kos97}. Hence, a study of the penetration
depth at low temperatures $T_{s}\ll T\sim T_{m0}\ll T_c$ may be carried
out disregarding nonlocal effects. Then a magnetic field enters into
Eq.(\ref{gcur}) only together with the Matsubara frequencies $\varepsilon_n-
i\frac{\displaystyle e}{\displaystyle c}{\rm v}_{f,y}A(x)$ in the argument of
the Green's function. The kernel $Q(x,T)$ can then be
written
\begin{equation}
Q(x,T)=\frac{\displaystyle 2ie^2 TN_f}{\displaystyle c}\sum_{n=-\infty}^{
+\infty}
\Big<{\rm v}_{f,y}^2(\bbox{p}_f)\frac{\displaystyle \partial g(\bbox{p}_f,x,
\varepsilon_n)}{\displaystyle\partial \varepsilon_n} \Big>_{S_f} \ .
\label{kernel}
\end{equation}

In the presence of zero-energy surface bound states, the pole-like term in the
propagator becomes dominating at temperatures $T\ll T_c$. Surface
bound states as well as their paramagnetic response are localized
on the scale of the coherence length at the surface, however, while the
conventional screening current has an avenue of the huge
thickness of the penetration depth. That is why the
zero-energy bound-state contribution to the penetration depth remains a
small low-temperature correction to $\lambda_0\equiv\lambda_0(T=0)$
at all temperatures $T\gg T_{s}$ (in particular, at $T\sim T_{m0}$). The
contribution from surface bound states must be viewed together with a low
temperature correction from the screening current as small
low-temperature imports to the zero temperature London penetration depth
$\lambda_0$. Then the total kernel of the form $Q(x,T)=\frac{
\displaystyle c}{\displaystyle4\pi\lambda_0^2}+\delta Q(x,T)$ includes
only the lowest order corrections in $\delta Q(x,T)$.

Solving the Maxwell equation
\begin{equation}
A''(x)-\frac{\displaystyle 1}{\displaystyle \lambda_0^2}A(x)-\frac{
\displaystyle 4\pi}{\displaystyle c}\delta Q(x,T)A(x)=0
\label{max}
\end{equation}
perturbatively with respect to the last term delivers a first order
approximation to the vector potential:
\begin{equation}
A(x)=A^{(0)}(0)\left[\exp\left(-\frac{\displaystyle x}{\displaystyle \lambda_0}
\right)-\frac{\displaystyle 2\pi\lambda_0}{\displaystyle c}\int\limits_0^{+
\infty}dx'\exp\left(-\frac{\displaystyle |x-x'|}{\displaystyle\lambda_0 }
\right)\delta Q(x',T)\exp\left(-\frac{\displaystyle x'}{\displaystyle\lambda_0}
\right)\right]\enspace .
\label{a}
\end{equation}
The kernel $\delta Q(x,T)$ incorporates only a contribution from the
bound states and a low-temperature correction from the screening current.

The penetration depth is defined as $\lambda=\int_0^{+\infty} H(x)dx/H(0)=
-A(0)/A'(0)$. Expanding this to first order in $\delta Q$ and extracting the
low temperature correction from the screening current for the case of a
superconductor with a line of nodes
\begin{equation}
\label{lambda902}
\lambda(T)=\lambda_0+a\lambda_0 \frac{\displaystyle T}{\displaystyle T_c}
-\frac{\displaystyle 4 \pi \lambda_0^2}{\displaystyle c } \int\limits_{0}^{
\infty }Q^{bound}(x,T) dx
\enspace .
\end{equation}
Here $a$ is a coefficient of the order of unity which depends on the shape
of the Fermi surface and on an angular slope of the order parameter near
the nodes. For instance, for a quasi-two-dimensional $d_{x^2-y^2}$ tetragonal
superconductor with a cylindrical Fermi surface (with a principal axis $z$) and
order parameter $\Delta(\phi)=\Delta_0\cos(2\phi-2\alpha)$,
one gets $a\approx0.32$.

Kernel $Q^{bound}(x,T)$ takes negative values. It is a paramagnetic
contribution from zero-energy bound states to Eq.(\ref{kernel}). One obtains
$Q^{bound}(x,T)$ from Eq.(\ref{kernel}) substituting instead of the full
expression for $g(\bbox{p}_f,x,\varepsilon_n)$ only its singular part
(pole-like term) $g_s(\bbox{p}_f,x,\varepsilon_n)$. Associated with zero energy
surface bound states, this term vanishes in the bulk on the scale of the
coherence length $\xi_0$. It has longer tails only towards the nodes. Node
contributions do not dominate, however, in the following expressions. The
presence of zero-energy surface bound states is crucial in the reasoning. All
sectors of the Fermi surface associated with a sign change of the order
parameter in a quasiparticle reflection from the surface, contribute
significantly to the results. This allows us to neglect, to a good accuracy,
the factor $\exp(-2x/\lambda_0)$ under the integral sign in
Eq.(\ref{lambda902}).

The analytic expression for the pole-like term has been found in the clean
limit and for a smooth surface in Ref.\ \onlinecite{bs97}:
\begin{equation}
\label{gs1}
g_s(\bbox{p}_f,x,\varepsilon_n)=\frac{\displaystyle -2\pi i}{\displaystyle
\varepsilon_n}\frac{\displaystyle|\tilde \Delta (\bbox{p}_f,0)||\tilde\Delta(
\underline{\bbox{ p}}_f,0)|}{\displaystyle |\tilde\Delta
(\bbox{p}_f,0)+|\tilde\Delta(\underline{\bbox{p}}_f,0)|}
\Theta(\bbox{p}_f)\exp{\left(-\frac{\displaystyle 2}{\displaystyle
|{\rm v}_{f,x}(\bbox{p}_f)|}
\int_0^x |\Delta({\bf p}_f,x^{\prime })|dx^{\prime }\right)}
\enspace .
\end{equation}

The effective surface order parameter $|\tilde{\Delta}(\bbox{p}_{f},0)|$
introduced in Eq.(\ref{gs1}), is defined
\begin{equation}
\frac{\displaystyle 1}{\displaystyle |\tilde{\Delta}({\bbox{p}}_{f},0)|}=
\frac{\displaystyle 2}{\displaystyle |{\rm v}_{f,x}(\bbox{p}_{f})|}
\int\limits_0^{\infty}
\exp\left(-\displaystyle\frac{\displaystyle 2}{\displaystyle |{\rm
v}_{f,x}
(\bbox{p}_{f})|}\int\limits_0^{x}
|\Delta(\bbox{p}_{f},
x')|dx' \right)dx \enspace .
\label{|effdel|}
\end{equation}

Here we distinguish between incoming $\bbox{p}_f$ and outgoing
$\underline{{\bbox{p}}}_f$ quasiparticle momenta in a reflection event. For
specular reflection, the momentum parallel to the interface is conserved.
Function $\Theta(\bbox{p}_f)$ is equal to unity  where zero energy bound states
occur on the Fermi surface (i.e. where the order parameter in the bulk taken
for incoming $\bbox{p}_f$ and outgoing $\underline{\bbox{p}}_f$ momentum
directions have opposite signs), and vanishes elsewhere.

Substituting Eq.(\ref{gs1}) in Eq.(\ref{kernel}), one can easily sum
over the Matsubara frequencies.
Integration over the space coordinate $x$ in
Eq.(\ref{lambda902}) then yields the penetration depth:
\begin{equation}
\label{lambda8}
\lambda(T)=\lambda_0 +a\frac{\displaystyle T}{\displaystyle T_c}\lambda_0 +
\frac{\displaystyle\pi^2e^2N_f\lambda_0^2}{\displaystyle c^2T}\left<
{\rm v}_{f,y}^2(\bbox{p}_f)|{\rm v}_{f,x}(\bbox{p}_f)|
\Theta(\bbox{p}_f)\right>_{S_f}
\enspace ,\qquad
T_c\frac{\displaystyle \xi_0}{\displaystyle\lambda_0}\ll T\ll T_c \enspace .
\end{equation}
For a three dimensional superconductor with a spherical Fermi surface one has
the relation $\lambda^2_0=3c^2/(8\pi e^2{\rm v}_f^2N_f)$. Then the coefficient
in front of the third term in Eq.(\ref{lambda8}) is $\frac{\displaystyle
3\pi}{\displaystyle 8T{\rm v}_f^2}$. Analogously, for a simple model of a
quasi-two-dimensional superconductor with a cylindrical Fermi surface,
$\lambda^2_0=c^2/(4\pi e^2{\rm v}_f^2N_f)$ and the coefficient
$\frac{\displaystyle \pi}{\displaystyle 4T{\rm v}_f^2}$.

In particular, for a $d_{x^2-y^2}$-wave superconductor with a cylindrical
Fermi surface, we get from Eq.(\ref{lambda8})
\begin{equation}
\label{lambda9}
\lambda(T)=\lambda_0 +a\frac{\displaystyle T}{\displaystyle T_c}\lambda_0 +
\frac{\displaystyle {\rm v}_f}{\displaystyle 6T}\left| |\sin^3\beta|-
|cos^3\beta|\right|
\enspace,\qquad
T_c\frac{\displaystyle \xi_0}{\displaystyle\lambda_0}\ll T\ll T_c \enspace ,
\end{equation}
where $\beta=\alpha+(\pi/4)$ is the angle between the surface normal and the
direction to a node of the order parameter, while $\alpha$ is the angle between
the surface normal and the crystalline $a$-axis along its positive lobe.

We note that the correction from zero energy bound states to the penetration
depth (the third term in Eq.(\ref{lambda8})) has a quite universal form. It is
independent both of the spatial profile of the order parameter near a surface
and its particular anisotropic structure (basis functions). Therefore, this
correction depends only on the type of pairing, which determines regions on the
Fermi surface with opposite signs of the order parameter. For example,
expression (\ref{lambda9}) is valid irrespective of a particular form of a
momentum direction dependence of the basis function for a $d$-wave order
parameter of given symmetry.

The ratio of a supercurrent density at the
surface $j^{bound}_s(x=0,T)$ to the one $j_s(x_{scr},T)$ at a characteristic
distance $x_{scr}$ ($\xi_0\ll x_{scr}\ll \lambda_0$) from the surface can be
estimated for a clean superconductor\cite{clean} at $T\ll T_c$ and a smooth
surface as $|j^{bound}_s(x=0,T)/ j_s(x_{scr},T)|\sim 4\pi\lambda_0^2|Q^{bound}
(x=0,T)|/c \sim T_c/T$. This verifies that at low temperatures $T\ll T_c$ the
paramagnetic current $j^{bound}_s(x,T)$ dominates over the shielding current
near the surface within a relatively small characteristic scale $\xi_0$.

The temperature dependent terms in Eq.(\ref{lambda8}) behave in very
different fashions from each other. They come from the conventional
shielding currents and from the zero-energy bound states. Growing
with decreasing temperature, the diamagnetic screening currents monotonically
reduce the penetration depth. On the other hand, Andreev surface-bound states
respond paramagnetically and increase
the penetration depth when the temperature goes down. Disregarding the
broadening effects, Eq.(\ref{lambda8}) delivers the following estimate for
the field of the low-temperature minimum of the penetration depth:
\begin{equation}
T_{m0}=\zeta\sqrt{\frac{\displaystyle \xi_0}{\displaystyle \lambda_0}}T_c
\label{min}\enspace ,
\end{equation}
where $\zeta$ is of the order of unity for crystalline orientations with
sufficient amount of momentum directions admitting zero-energy bound states.
Otherwise $\zeta$ is a small quantity. For a $d$-wave superconductor $\zeta
\propto\left| |\sin^3\beta|-|cos^3\beta|\right|^{1/2}$ and vanishes for
$\beta=45^\circ$ (i.e. for $\alpha=0$), when there are no zero energy bound
states.

Broadening of the bound states can substantially modify the
conditions for the presence of a minimum in the low temperature
dependence of the penetration depth.  For a small broadening
$\gamma(\bbox{p}_f)\ll T_c$ we simply replace the factor
$\frac{\displaystyle 1}{\displaystyle \varepsilon_n}$ in the
expression Eq.(\ref{gs1}) for the pole-like term with
$\frac{\displaystyle 1}{\displaystyle
\left[\varepsilon_n+\gamma(\bbox{p}_f){\rm sgn}(
\varepsilon_n)\right]}$. Taking into account the broadening
Eq.(\ref{lambda8}) is generalized to the following form:
\begin{equation}
\label{lambda1}
\lambda(T)=\lambda_0 +a\frac{\displaystyle T}{\displaystyle T_c}\lambda_0 +
\frac{\displaystyle 2e^2N_f\lambda_0^2}{\displaystyle c^2T}\left<
{\rm v}_{f,y}^2(\bbox{p}_f)|{\rm v}_{f,x}(\bbox{p}_f)|
\Theta(\bbox{p}_f)\psi'\left(\frac{\displaystyle1}{\displaystyle2}
+\frac{\displaystyle\gamma(\bbox{p}_f)}{\displaystyle2\pi T }\right)
\right>_{S_f} \, .
\end{equation}
Here and below $\psi(x)$ is the digamma function and $\psi'(x)$ - its
derivative.

Eq.(\ref{lambda1}) is a reasonable representation of the role of a broadening
in the low temperature anomaly of the penetration depth. The minimum lies
at $T_{m0}\approx1.8 \sqrt{\xi_0/\lambda_0}T_c$  for momentum independent
broadening in a $d_{x^2-y^2}$-superconductor in the clean limit
$\gamma\ll\pi T$ with the orientation $\alpha=45^\circ$. With increasing
broadening it drifts to lower
temperatures (becoming less pronounced at the same time) till $T_{m\gamma}
\approx0.4 \sqrt{\xi_0/\lambda_0}T_c$ at $\gamma\approx0.96T_{m0}$, where it
evaporates. As an example, the low-temperature correction to the penetration
depth is shown in Fig.\ 1 in the vicinity of $T_{m\gamma}$ for various values
of the momentum independent broadening.

\begin{figure}[h]
\centerline{
\psfig{figure=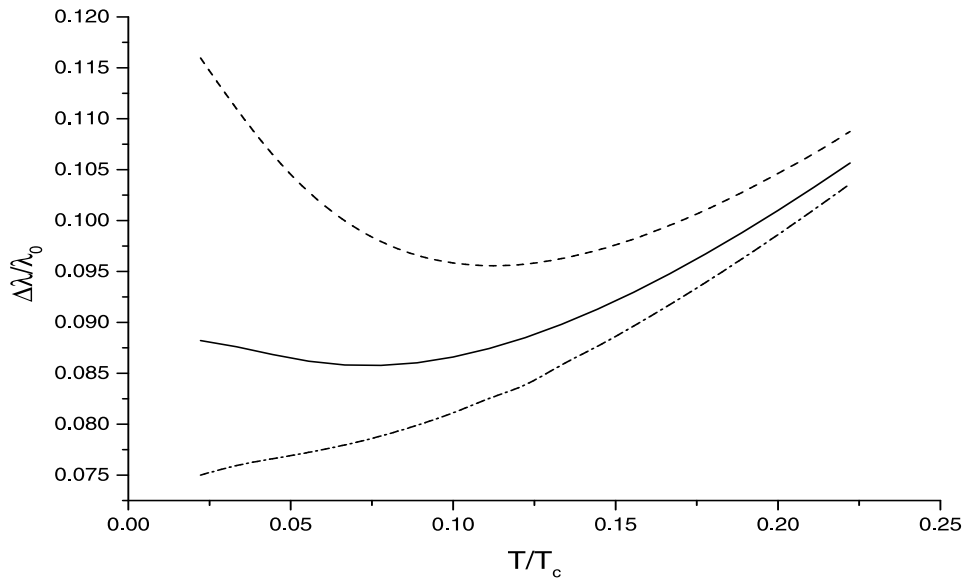,width=5in,height=3in}
\label{RSh}
}
\vspace{0.3cm}
\caption{ Low-temperature correction to the penetration depth (in units of
$\lambda_0$) in a $d_{x^2-y^2}$ superconductor with a cylindrical Fermi
surface and the orientation $\alpha=45^\circ$. The temperature is measured in
units of $T_c$. The parameter
$\xi_0/\lambda_0$ is chosen to be $\approx 0.01$, where $\xi_0=({\rm v}_f/
2\pi T_c)$. The curves are given for three values of the broadening: $\gamma=
0.10T_c$ (dashed line), $\gamma=0.15T_c$ (solid line) and $\gamma=0.19T_c$
(dashed-dotted line).
}
\end{figure}

There are various contributions to the broadening of the bound states
associated, in particular, with surface roughness, nonmagnetic and magnetic
impurities and inelastic scattering. We now pin-point the origin of the
broadening, assuming that nonmagnetic impurities dominate the scene.

With Born scatterers $\gamma_b\approx \sqrt{\frac{\displaystyle T_c}{
\displaystyle\tau}}$ (see Ref.\ \onlinecite{pbbi99}) and the coefficient of the
order of unity can be estimated within the simple model of spatially constant
order parameter. Then we easily get the shortest normal-state impurity mean
free path $l$ which admits a low-temperature upturn: $\lambda_0\lesssim l$.
In high-temperature superconductors one should distinguish between $l$ and
the actual mean free path in the normal state at $T_c$ incorporating
significant contributions from inelastic processes. Impurity scattering
dominates there at low temperatures already in the superconducting state
where the collapse of inelastic scattering takes place. For instance,
below 20 K in $YBa_2Cu_3O_{7-\delta}$ there is a regime of extremely long
and weakly temperature dependent quasiparticle scattering
times\cite{bonn92,bonn94,krish95,krish99,hosseini99,berlinsky00} usually
interpreted as due to feeble impurity scattering in high-purity samples.

For scatterers with sufficient strength of impurity potential there are
practically no restrictions on the impurity scattering rate in contrast to
what was found above for Born impurities. For unitary scatterers with
scattering rates $\Gamma_u\ll T_c$ the broadening of the zero-energy bound
states is exponentially small\cite{pbbi99}:
$\gamma_u=B\sqrt{\Delta_0\Gamma_u}\exp\left(-b\Delta_0/ \Gamma_u\right)$. A
scattering rate $\Gamma_u$ which leads to a given broadening $\gamma_u$ is
almost independent of a constant coefficient $B$ in the pre-exponential factor,
while it is sensitive to the model dependent parameter $b$ in the argument of
the exponential function. Within the simple model considered in
Ref.\ \onlinecite{pbbi99}, one gets $b\sim 1$.

For temperatures $T\lesssim\sqrt{\Gamma_u\Delta_0}$, the share of the
penetration depth from the shielding currents must be modified for unitary
scatterers. This leads instead of the linear term in Eqs.(\ref{lambda902}),
(\ref{lambda8}), (\ref{lambda9}), (\ref{lambda1}) to a quadratic low
temperature correction of the form
$\lambda_0T^2/\left(\Gamma^{1/2}\Delta_0^{3/2}\right)$ to within a factor the
order of unity\cite{hir93,bonn94}. Correspondingly, $T_{m0}$ given in
Eq.(\ref{min}) is valid for unitary scatterers only if
$T_{m0}>\sqrt{\Gamma_u\Delta_0}$, which sets an upper limit on the scattering
rate: $\Gamma_u<\frac{\displaystyle \xi_0}{\displaystyle\lambda_0}\Delta_0$.

The $T^2$-term instead of the linear one in Eq.(\ref{lambda8}) delivers
an estimate for the location $T_{md}$ of the low temperature
minimum of the penetration depth modified by unitary scatterers:
\begin{equation}
T_{md}\approx\left(\frac{\displaystyle\xi_0}{\displaystyle
\lambda_0}\right)^{1/3}\left(\Gamma_u\Delta_0^5\right)^{1/6}
\enspace, \qquad  \Gamma_u>\frac{\displaystyle\xi_0}{\displaystyle\lambda_0}
\Delta_0\enspace .
\label{unlim}
\end{equation}

This expression replaces Eq.(\ref{min}) in the case of unitary scatterers with
the scattering rate $\Gamma_u>\frac{\displaystyle\xi_0}{\displaystyle\lambda_0}
\Delta_0$. The minimum slowly drifts to higher temperatures with
increasing $\Gamma_u$. It does not melt away at any
$\Gamma_u\ll T_c$. The normal-state impurity mean free path must just be large
on the scale of the coherence length.

We conclude that observation of the low temperature upturn
of the penetration depth in samples with $l<\lambda_0$ is evidence for both
Andreev bound states and a sufficiently large strength of the bulk impurity
potential in the superconducting compounds. For unitary impurities one
needs to take into account the broadening that arises from surface roughness
which then very probably controls the total broadening. The
same effect with Born scatterers demands the normal-state impurity mean free
path larger than the London penetration depth.

\section{Zero energy bound states and spontaneous surface current}

Throughout this section the broadening of the zero-energy bound
states is assumed small.
We look at a clean $d$-wave superconductor with a smooth surface.
Its crystal-to-surface orientation shall admit zero-energy surface
bound states and feature an upturn in the penetration depth.
Below the upturn temperature $T_{m0}$, imagine a great deal
of space for $\lambda$ to grow, firstly as described by the perturbative result
Eq.(\ref{lambda8}). Then a second order phase transition occurs
at $T\sim T_{s}\ll T_{m0}$ into a state which carries a
spontaneous surface supercurrent\cite{hig97,sigr99,belz99}. We
shall find an analytic expression for the transition temperature
and discuss the impact of impurities on the effect. The transition implies
the absence of subdominant channels activated at low temperatures
close to the surface on account of the presumably large surface pair breaking
in the dominant component of the order parameter.
Otherwise a spontaneous current can arise at higher
temperatures\cite{buch95,mats95,fog97,sigr99}.

There is experimental evidence \cite{cov97} for a phase transition on the (110)
surface in $YBa_2Cu_3O_{7-\delta}$ at $T=7K\gg(\xi_0/\lambda_0)T_c$. It was
interpreted as associated with an activated near surface subdominant channel of
the order parameter\cite{fog97}. For some other crystal-to-surface
orientations, however, a subdominant component can be not present near a
surface\cite{buch95,mats95}. Zero energy bound states can still arise for a
noticeable part of quasiparticle trajectories. Our theoretical study is
relevant to these cases.

In order to find an equation for the transition temperature, one has to
admit a paramagnetic contribution to the penetration depth at least as
large as the diamagnetic one. Then a perturbation treatment of the preceding
section is not adequate. In this context we develop an approach based on
the integral form of Eq.(\ref{max}) and take into account only the terms in
$\delta Q(x,T)$ brought about by the bound states. In other words, a
contribution only from the pole-like term Eq.(\ref{gs1}) needs to be taken into
account in Eqs.(\ref{kernel}) for the kernel which enters into Eq.(\ref{max}).
The kernel $\delta Q(x,T)$ varies on the characteristic scale $\xi_0$ and is
associated in the clean limit\cite{clean} with large contributions to the
magnetic field at the surface at temperatures $T\lesssim (\xi_0/\lambda_0)T_c$.
We therefore disregard the nonlocal temperature correction from the Meissner
current to $\delta Q(x,T)$.

We transform Eq.(\ref{max}) into the integral form
$$
A(x)=\left[A(0)-\frac{\displaystyle 2\pi\lambda_0}{\displaystyle c}
\int\limits_0^{+\infty}dx'Q^{bound}(x',T)A(x')\left(e^{\frac{\displaystyle
x'}{\displaystyle\lambda_0}}-e^{-\frac{\displaystyle x'}{
\displaystyle\lambda_0}}\right)\right]
e^{-\frac{\displaystyle x}{
\displaystyle\lambda_0}} -
\qquad \qquad \qquad \qquad
$$
\begin{equation}
\qquad  \qquad \qquad \qquad \qquad \qquad
-\frac{\displaystyle 2\pi\lambda_0}{\displaystyle c}
\int\limits_x^{+\infty}dx'Q^{bound}(x',T)A(x')\left(e^{
\frac{\displaystyle x-x'}{\displaystyle \lambda_0}}-e^{
-\frac{\displaystyle x-x'}{\displaystyle \lambda_0}}\right) \enspace .
\label{integ}
\end{equation}

The two terms on the right hand side of this equation obey very different
scales. The first decays exponentially in the depth on
the scale $\lambda_0$ while the last term vanishes for $x\gg \xi_0$ along
with the kernel $Q^{bound}(x,T)$.
The kernel $Q^{bound}(x=0,T)$  can be estimated (see preceding section)
for a clean
superconductor and a smooth surface as $2\pi\lambda^2_0Q^{bound}( x=0,T)/c\sim
T_c/T$. Then, in accordance with Eq.(\ref{integ}), the approximate formula
$\left(1-\frac{\displaystyle A(x_{scr})}{\displaystyle A(0)}\right)\sim\xi^2_0
T_c/\lambda_0^2T$ is established for a relative deviation of the vector
potential $A(x_{scr})$ taken at the distance $x_{scr}$ ($\xi\ll
x_{scr}\ll \lambda_0$) from its value $A(0)$ at the surface. The deviation
reflecting the bound state contribution to the vector potential turns out
to be small at all temperatures $T\gg \left(\xi^2_0/\lambda_0^2\right)T_c$,
in particular, for $T\sim T_s\sim\left(\xi_0/\lambda_0\right)T_c$.
Varying on the scale $\xi_0$, small terms in the expression for the vector
potential at temperatures $T\sim T_s$ are of importance only when
differentiating $A(x)$. After that they can already noticeably contribute
to the expression for the magnetic field.

Indeed, a spatial differentiation of Eq.(\ref{integ}) leads to
$$
H(x)=-\frac{\displaystyle 1}{\displaystyle \lambda_0}\left[A(0)-
\frac{\displaystyle 2\pi\lambda_0}{\displaystyle c}
\int\limits_0^{+\infty}dx'Q^{bound}(x',T)A(x')\left(e^{\frac{\displaystyle
x'}{\displaystyle\lambda_0}}-e^{-\frac{\displaystyle x'}{
\displaystyle\lambda_0}}\right)\right]
e^{-\frac{\displaystyle x}{
\displaystyle\lambda_0}} -
\qquad \qquad \qquad
$$
\begin{equation}
\qquad  \qquad \qquad \qquad \qquad \qquad
-\frac{\displaystyle 2\pi}{\displaystyle c}
\int\limits_x^{+\infty}dx'Q^{bound}(x',T)A(x')\left(e^{
\frac{\displaystyle x-x'}{\displaystyle \lambda_0}}+e^{
-\frac{\displaystyle x-x'}{\displaystyle \lambda_0}}\right) \enspace .
\label{integ2}
\end{equation}

The second term in the square brackets remains negligibly small $\sim\left(
\xi^2_0T_c/\lambda_0^2T\right) A(0)\ll A(0)$ as compared with $A(0)$ for $T\gg
\left(\xi^2_0/\lambda_0^2\right)T_c$. The last term of
Eq.(\ref{integ2}) is the order of $\left(\xi_0T_c/\lambda_0T\right) (A(0)/
\lambda_0)$. For a deviation of the magnetic field at $x=x_{scr}$ from its
value at $x=0$:\ $\left(\frac{\displaystyle H(x_{scr})}{\displaystyle H(0)}-
1\right)\sim\left(\xi_0 T_c/\lambda_0 T\right)$. Hence, the bound state
contribution to the magnetic field can be viewed as a small perturbation as
compared with the shielding contribution unless $T\lesssim T_s$. Considering
$\left(\xi^2_0/ \lambda_0^2\right)T_c\ll T \lesssim T_s$, we can discard the
second term in the square brackets but have to keep track of the last term in
Eq.(\ref{integ2}). Choosing $x=0$ in Eq.(\ref{integ2}), the small terms
$x'/\lambda_0$ in the exponential functions under the integral sign can be
taken to vanish. For the same reason and within the same accuracy one can
treat the vector potential under the integral sign in Eq.(\ref{integ2}) as
constant in space $A(0)$ discarding small terms in the vector
potential which vary on the scale $\xi_0$. All this results in an explicit
relation between $A(0)$ and $H(0)$ and therefore
\begin{equation}
\lambda=\frac{\displaystyle\lambda_0}{\displaystyle 1+\frac{\displaystyle 4
\pi\lambda_0}{\displaystyle c}\int\limits_0^{+\infty}Q^{bound}(x,T)dx}
\enspace .
\label{sing}
\end{equation}

Proceeding like in the derivation of Eq.(\ref{lambda8}) above,
we find that the paramagnetic (negative) sign of $Q^{bound}$
leads to a divergence of $\lambda$ at the temperature
\begin{equation}
T_s=\frac{\displaystyle \pi^2e^2N_f\lambda_0}{\displaystyle c^2}\left<
{\rm v}_{f,y}^2(\bbox{p}_f)|{\rm v}_{f,x}(\bbox{p}_f)|
\Theta(\bbox{p}_f)\right>_{S_f}
\enspace .
\label{ts}
\end{equation}
For the model $d$-wave superconductor with  a cylindrical Fermi surface
one gets from Eq.(\ref{ts})
\begin{equation}
T_s=\frac{\displaystyle \pi\xi_0}{\displaystyle 3\lambda_0}\left||\sin^3\beta|
-|cos^3\beta|\right|T_c
\enspace ,
\label{dts}
\end{equation}
where $\xi_0={\rm v}_f/2\pi T_c$.

The divergence of $\lambda$ implies the existence of a nontrivial solution
to Eq.(\ref{integ}) in a vanishing external magnetic field. Indeed, if we let
$H(0)=0$, $A(0)\ne0$, then Eq.(\ref{integ2}) transforms,
with the same approximation as above, into the relation:
$1=-\left(4\pi\lambda_0/c\right)\int_0^{+\infty} Q^{bound}(x',T)dx'$,
which defines the transition temperature $T_s$ into a state with a spontaneous
surface supercurrent.

The nontrivial solution at $T_s$ is a result of interplay between the
paramagnetic supercurrent which originates in the zero energy bound states
localized within $\xi_0$ on the one hand and the diamagnetic supercurrent
distributed over the region $x\sim\lambda_0$ on the other. The latter
compensates for the magnetic field from the bound states at the surface in
order to satisfy the boundary conditions in the absence of an external magnetic
field. Then $\int\limits_0^{+\infty}j(x)dx=0$ always applies being a
consequence of the full screening of the spontaneous surface magnetic field in
the bulk of a superconductor. Under these conditions the Bloch theorem, in
general, admits spontaneous surface currents\cite{ohashi96}.
The magnetic part of a superconducting free energy $\frac{\displaystyle
1}{\displaystyle8\pi}\int\limits_0^{+\infty} \biggl[A'^2(x)+\frac{\displaystyle
4\pi}{\displaystyle c}Q(x,T)A^2(x)\biggr]dx$ vanishes at $T=T_s$ and
becomes negative below $T_s$ on account
of negative sign of the paramagnetic kernel $Q^{bound}$ (Gibbs and Helmholtz
free energies coincide in zero external magnetic field). The
result is an energetically favorable state with a spontaneous surface
supercurrent below $T_s$.

The broadening $\gamma$ of the bound states modifies Eq.(\ref{ts}):
\begin{equation}
T_s=\frac{\displaystyle 2e^2N_f\lambda_0}{\displaystyle c^2}\left<
{\rm v}_{f,y}^2(\bbox{p}_f)|{\rm v}_{f,x}(\bbox{p}_f)|
\Theta(\bbox{p}_f)\psi'\left(\frac{\displaystyle1}{\displaystyle2}+\frac{
\displaystyle\gamma(\bbox{p}_f)}{\displaystyle2\pi T_s }\right)\right>_{S_f}
\enspace .
\label{tsgamma}
\end{equation}

The broadening prevents the appearance of a spontaneous
surface current unless $\gamma\lesssim\frac{\displaystyle\xi_0}{\displaystyle
\lambda_0} T_c$. This is a very strong restriction. If Born
impurities control the broadening, they admit spontaneous surface
supercurrent only with extremely large values of the mean free path
$\lambda^2_0/\xi_0\sim100\lambda_0 \lesssim l$, unrealistic for high
temperature
superconductors. Unitary scatterers impose a much weaker restriction
$\Gamma_u\lesssim2b\Delta_0 \left/\ln\left[\frac{\displaystyle\lambda^2_0}{
\displaystyle\xi^2_0\ln(\lambda_0/\xi_0)}\right]\right.\sim 0.1T_c$. Then,
however, surface roughness probably dominates the broadening and can
destroy the state with a spontaneous surface supercurrent.

\section{Nonlinear Meissner effect from low energy bound states}

It is important in the derivation of Eq.(\ref{sing}) that the kernel
$Q^{bound}$ varies much faster in space than the screening currents. Then
contributions of the paramagnetic current carried by surface Andreev-bound
states at temperatures $(\xi_0/\lambda_{0})^2T_c\ll T$, can result in
significant spatial variations of the magnetic field near the surface while in
the weakly spatially dependent vector potential. This leads to Eq.(\ref{sing})
on the basis of the local current-field relation.

It is straightforward to show within the same framework that a nonlinear
penetration depth $\lambda_{nl}(T,H)$ incorporating contributions both from
screening currents and from zero-energy bound states is described as
\begin{equation}
\lambda_{nl}(T,H)=\frac{\displaystyle\lambda^{scr}_{nl}(T,H_{scr})}{
\displaystyle 1+ \frac{\displaystyle 4\pi\lambda^{scr}_{nl}(T,H_{scr})}{
\displaystyle c}\int\limits_0^{+\infty}Q^{bound}_{nl}\Bigl(x,T\Bigr)dx}
\enspace ,
\label{sing2}
\end{equation}
where $\lambda^{scr}_{nl}(T,H_{scr})$ is a contribution from screening
supercurrents to $\lambda_{nl}(T,H)$, taken at an effective value of the
field $H_{scr}=H(0)-\frac{\displaystyle4\pi}{\displaystyle
c}\!\int\limits_0^{+\infty}\!H(x)dx\int\limits_0^{+\infty}\!Q^{bound}_{nl}
\Bigl(x',T\Bigr)dx'$. Here $H(0)$ is the external magnetic field.
The second term describes the field of the zero-energy bound states inside the
superconductor at distances $x\approx x_{scr}$ ($\xi_0\ll x_{scr}\ll
\lambda_0$), as can be seen in Eq.(\ref{integ2}). A paramagnetic response of
zero-energy bound states ($Q^{bound}_{nl}<0$) increases the field to be
screened by diamagnetic supercurrents ($H_{scr}\equiv H(x_{scr})>H(0)$).
This leads, in general, to more pronounced nonlinear terms in
$\lambda^{scr}_{nl}(T,H_{scr})$ as compared to disregarding the contribution
from zero-energy bound states. In the case of spontaneous surface supercurrent
$H_{scr}$ differs from zero even in the absence of an external field. We assume
the condition $H_{scr}<H_{c1}$ for the Meissner state to be stable in the
magnetic field on account of a paramagnetic influence of the bound states.

Small nonlocal low
temperature corrections to the penetration depth from screening
currents can be taken into account in Eq.(\ref{sing2})
as perturbations to $\lambda^{scr}_{nl}(T,H)$. For
a nonlocal current-field relation a penetration depth
$\lambda_{nl}(T,H)$ is actually a functional of the spatial profile
of the magnetic field.

Nonlinear corrections from the shielding supercurrent to the
Meissner effect can be given in terms of the dimensionless ratio
$\rho=(H/H_0)$, where $H_0$ is usually the order of the
thermodynamic critical field $\sim \Phi_0/(\lambda_0\xi_0)$. Hence, they are
always small in strong type II superconductors in the Meissner state.
In isotropic $s$-wave superconductors, the first nonlinear correction to the
penetration depth $\propto\rho^2$. In superconductors with nodes in the
order parameter (for instance, $d$-wave ) a term linear in $\rho$ can arise
for particular crystal orientations at low
temperatures\cite{sauls92,sauls95}. The linear term,
however, is quite sensitive to nonlocal effects\cite{hirschfeld00} and
the impurity influence, in particluar, at sufficiently large strength of
impurity potentials\cite{sauls95,scalapino99}.

A nonlinearity in the magnetic response of low energy Andreev surface bound
states has, in general, a very different field scale
$\widetilde{H}_0$. In a clean limit
$\widetilde{H}_0(T)=(\Phi_0T/\lambda v_f)$, where $\lambda$ is determined
by Eq.(\ref{sing2}). At low temperatures $T\ll T_c$ one
always gets $\widetilde{H}_0(T)\ll H_0$. For instance,
$\widetilde{H}_0(T_{m0})\sim\sqrt{\frac{\displaystyle\xi_0}{
\displaystyle\lambda_0}}H_0\sim0.1H_0$, $\widetilde{H}_0(T_{s})\sim
\frac{\displaystyle \lambda_0}{\displaystyle\lambda}H_{c1}\lesssim H_{c1}\sim
0.01 H_0$. Moreover, at sufficiently low temperatures $T\ll T_s$, i.e. well
below the transition to the state with a spontaneous surface supercurrent, a
paramagnetic response from the bound states may become seriously nonlinear
already in the Meissner state $\widetilde{H}_0(T)\sim H<
H_{c1}$\cite{hig97,belz99}. We will show, however, that the broadening of the
bound states introduces a new field scale
$\widetilde{H}_\gamma=\frac{\displaystyle\gamma\lambda_0}{ \displaystyle
T_c\lambda}H_0$ coming into play at $\pi T\lesssim\gamma$.  For
$\gamma\gg\frac{\displaystyle\xi_0}{\displaystyle\lambda_0}T_c\sim0.01T_c$
nonlinear corrections from Andreev low-energy bound states to the penetration
length turn out always to be small in the Meissner state, even at $T=0$.

As a pole-like term Eq.(\ref{gs1}) decays exponentially on the
scale $\sim\xi_0$ for almost all momentum directions admitting bound
states, we consider a local nonlinear current-field relation
\begin{equation}
\bbox{j}({x})=2eTN_f\sum\limits_{\varepsilon_n}\left<{\rm\bbox{v}}_fg_s(
\bbox{p}_f,{x},\varepsilon_n-i\frac{\displaystyle e}{\displaystyle c}
{\rm v}_{f,y}A(x))\right>_{S_f}
\enspace
\label{gacur}
\end{equation}
for the current via the bound states.
One can also set the vector potential $A(x)$ in the kernel
equal to $A(x=0)$. Then we easily generalize the reasoning
in the derivation of the third term in Eq.(\ref{lambda1}). Substituting
into Eq.(\ref{gacur}) the expression Eq.(\ref{gs1}) for the pole-like term with
the pole shifted in accordance with the broadening, we find:
\begin{equation}
\int\limits_{0}^{\infty }Q^{bound}_{nl}(x,T) dx =
\frac{\displaystyle ieN_f}{\displaystyle A(0)}\left<
{\rm v}_{f,y}(\bbox{p}_f)|{\rm v}_{f,x}(\bbox{p}_f)|
\Theta(\bbox{p}_f)\psi\left(\frac{\displaystyle1}{\displaystyle2}+\frac{
\displaystyle\gamma(\bbox{p}_f)+i\frac{\displaystyle e}{\displaystyle c}
{\rm v}_{f,y}(\bbox{p}_f)A(0)}{\displaystyle2\pi T }\right)\right>_{S_f}
\enspace .
\label{nl}
\end{equation}
In Eq.(\ref{nl}) the Fermi surface is assumed symmetric in reflections
across the $xz$-plane. Then averages over the Fermi surface of odd
powers of ${\rm v}_{f,y}$ vanish, no matter whether they are
multiplied by $|{\rm v}_{f,x}(\bbox{p}_f)|\Theta(\bbox{p}_f)$ or not. This
applies, in particular, to a $d$-wave superconductor with a cylindrical Fermi
surface whose principal axis $z$ is parallel to the boundary for arbitrary
orientations of the two other crystal axes $x_0$, $y_0$.

If $\frac{\displaystyle e}{\displaystyle c}{\rm v}_{f}A(0)\ll {\rm max}\bigl(
2\pi T, \gamma\bigr)$, one can expand the $\psi$-function in Eq.(\ref{nl}) in
powers of the small parameter ${\rm min}\biggl(H(0)\Big/\widetilde{H}_0(T),
H(0)\Big/\widetilde{H}_{\gamma}\biggr)$. Considering nonlinear corrections to
the penetration depth from screening currents $\Delta\lambda^{scr}_{nl}$
and bound states $\Delta\lambda^b_{nl}$ to be small, one can
represent them in the first approximation as additive contributions to
the total penetration depth $\lambda_{nl}(T,H)\approx \lambda(T)+
\Delta\lambda^{scr}_{nl}+\Delta\lambda^b_{nl}$.
The nonlinear correction from screening currents takes
the form $\Delta\lambda^{scr}_{nl}=\frac{\displaystyle\lambda^2(T)}{
\displaystyle\lambda_{0}^2(T)}\left[\lambda^{scr}_{nl}\left(T,H(0)\frac{
\displaystyle\lambda(T)}{\displaystyle\lambda_{0}(T)}\right)-\lambda_{0}(T)
\right]$. Quantities $\lambda(T)$ and $\lambda_0(T)$ being the zero-field
values of $\lambda_{nl}(T,H)$, $\lambda^{scr}_{nl}(T,H)$ respectively, satisfy
Eq.(\ref{sing}). Bound states renormalize nonlinear response from screening
currents already in this approximation.
Thus, the explicit analysis of $\Delta\lambda^{scr}_{nl}$
can be done combining the results of the preceding section and Refs.\
\onlinecite{sauls95,hirschfeld00,scalapino99}. Apart from too close to the
transition temperature $T_s$, the nonlinear correction to the penetration depth
from the bound states is:
\begin{equation}
\Delta\lambda^b_{nl}\approx -
\frac{\displaystyle e^4\lambda^4(T)N_fH^2(0)}{\displaystyle 12\pi^2c^4T^3}
\left<{\rm v}^4_{f,y}(\bbox{p}_f)|{\rm v}_{f,x}(\bbox{p}_f)|
\Theta(\bbox{p}_f)\psi^{(3)}\left(\frac{\displaystyle1}{\displaystyle2}+\frac{
\displaystyle\gamma(\bbox{p}_f)}{\displaystyle2\pi T }\right)\right>_{S_f}
\enspace .
\end{equation}

In the limit $\frac{\displaystyle e}{\displaystyle c}{\rm v}_{f}A(0)
\gg 2\pi T$ when the argument of the $\psi$-function in Eq.(\ref{nl}) is
large, we obtain
\begin{equation}
\int\limits_{0}^{\infty }Q^{bound}_{nl}(x,T) dx =-
\frac{\displaystyle eN_f}{\displaystyle A(0)}\left<
{\rm v}_{f,y}(\bbox{p}_f)|{\rm v}_{f,x}(\bbox{p}_f)|
\Theta(\bbox{p}_f)\arctan\left(\frac{\displaystyle e
{\rm v}_{f,y}(\bbox{p}_f)A(0)}{\displaystyle c\gamma(\bbox{p}_f)}
\right)\right>_{S_f}
\enspace .
\label{nl2}
\end{equation}
Then the broadening rather than the temperature fixes the bound
state contribution to the penetration depth.

As shown above, there is no state with a spontaneous surface current with
$\gamma\gg T_s$. Then $\lambda\sim \lambda_0$ and
$e{\rm v}_fA(0)\Big/c\gamma\ll H(0)/H_{c1}$. Since $H(0)<H_{c1}$
in the Meissner state we estimate $e{\rm v}_fA(0)\Big/c\gamma\ll 1$
and obtain in this limit from Eq.(\ref{nl2})
$$
\Delta\lambda^b_\gamma=
\frac{\displaystyle 4\pi e^2N_f\lambda_0^2}{\displaystyle c^2\gamma}\left<
{\rm v}_{f,y}^2(\bbox{p}_f)
|{\rm v}_{f,x}(\bbox{p}_f)|
\Theta(\bbox{p}_f)\right>_{S_f} -
\qquad \qquad  \qquad  \qquad  \qquad  \qquad  \qquad  \qquad  \qquad
$$
\begin{equation}
\qquad  \qquad  \qquad  \qquad  \qquad  \qquad  \qquad  \qquad
-\frac{\displaystyle 4\pi e^4\lambda_0^4N_fH^2(0)}{\displaystyle 3c^4\gamma^3}
\left<{\rm v}^4_{f,y}(\bbox{p}_f)|{\rm v}_{f,x}(\bbox{p}_f)|
\Theta(\bbox{p}_f)\right>_{S_f}
\, .
\label{lambdab}
\end{equation}

For a given $\lambda^{scr}_{nl}(T,H)$, Eq.(\ref{sing2}) in general should be
solved with respect to $\lambda_{nl}(T,H)$ in accordance with Eq.(\ref{nl}),
since $A(0)=-\lambda_{nl}(T,H)H(0)$. This is particularly important close
to the transition temperature $T_s$, where the Landau theory of
second order phase transitions is applicable. Then first nonlinear term turns
out to be the order of the zero-field paramagnetic contribution in the
denominator in Eq.(\ref{sing2}). Ignoring a weak field dependence of
$\lambda^{scr}_{nl}(T,H)$ stipulated by screening currents, we
obtain from Eqs.(\ref{sing2}), (\ref{nl}) the following equation for
$\lambda_{nl}(T,H)$:
\begin{equation}
\left(\frac{\displaystyle T}{\displaystyle T_s}-1\right)\kappa\lambda_{nl}(T,H)
+\eta H^2\lambda_{nl}^3(T,H)=\lambda^{scr}_{nl}(T_s)\enspace ,
\label{eql}
\end{equation}
where $H=H(0)$, $$\eta=\frac{\displaystyle e^4\lambda^{scr}_{nl}(T_s)N_f}{
\displaystyle 12\pi^2c^4T_s^3}\left<{\rm v}^4_{f,y}(\bbox{p}_f)|{\rm v}_{f,x}
(\bbox{p}_f)|\Theta(\bbox{p}_f)\psi^{(3)}\left(\frac{\displaystyle1}{
\displaystyle2}+\frac{\displaystyle\gamma(\bbox{p}_f)}{\displaystyle2\pi T_s}
\right)\right>_{S_f}\enspace, \enspace
\left. \kappa=1-\frac{\displaystyle
d{\tilde T}_s(T)}{\displaystyle dT}\right|_{T=T_s}\enspace .$$
$T_s$ is described by Eq.(\ref{tsgamma}) and $\tilde{T}_s(T)$ is the result of
the substitution $T_s\rightarrow T$ on the right hand side of
Eq.(\ref{tsgamma}). The broadening is assumed to be
sufficiently small for admitting the phase transition.

The role of the order parameter in the phase transition can be played
by a surface
magnetization $m_S=\!\int\limits_0^{+\infty}\left(M(x)-M_\infty\right)dx=
\frac{\displaystyle1}{\displaystyle c}\int\limits_0^{+\infty}dxxj_s(x)=
\frac{\displaystyle1}{\displaystyle
4\pi}\!\int\limits_0^{+\infty}\!H(x)dx=
\frac{\displaystyle1}{\displaystyle4\pi}\lambda_{nl}(T,H)H(0)$,
which, for simplicity, we choose constant in space along a smooth
surface. The magnetization ${\bf M}$ enters by the conventional
definition ${\bf j}=c\cdot{\rm curl}\,{\bf M}$, and $M_\infty=-H(0)/4\pi$.
Then the Landau free energy per unit surface ${\cal F}_S$ which leads
to the same equation for $m_S$ as implied in Eq.(\ref{eql}) has the form
\begin{equation}
{\cal F}_S= \tilde\alpha\cdot\left(\frac{\displaystyle T}{\displaystyle T_s}-1
\right)m_S^2+\tilde\beta m_S^4-m_SH\enspace ,
\end{equation}
where $\tilde\alpha=2\pi\kappa/\lambda^{scr}_{nl}(T_s)$, $\tilde\beta=16\pi^3
\eta/\lambda^{scr}_{nl}(T_s)$, $H$ is the external field. As for a conventional
order parameter in a strong field near $T_s$ one gets $m_S(T_s,H)\propto
H^{1/3}$, which entails $\lambda_{nl}(T_s,H)\propto H^{-2/3}$.

Finally, in the limit of very small broadening $\gamma \ll \frac{\displaystyle
H(0)}{\displaystyle H_0}T_c$, Eqs.(\ref{sing2}) and (\ref{nl2}) give
\begin{equation}
\Delta\lambda^b_H=\lambda_{nl}(T,H)-\lambda^{scr}_{nl}(T,H)=
\frac{\displaystyle 4\pi eN_f\lambda_{0}}{\displaystyle c |H(0)|}
\left<|{\rm v}_{f,y}(\bbox{p}_f){\rm v}_{f,x}(\bbox{p}_f)|\Theta(\bbox{p}_f)
\right>_{S_f}
\label{field}
\end{equation}
at temperatures
$\frac{\displaystyle \xi_0^2}{\displaystyle \lambda_0^2}T_c\ll T \ll
\frac{\displaystyle H(0)}{\displaystyle H_0}T_c$. Since $\Delta\lambda^b_H$
is at least the order of $\lambda_0$, we put here $\lambda^{scr}_{nl}(T,H)
\approx\lambda_{0}$ disregarding small nonlinear corrections from
the screening currents in the Meissner state.
The approximate inverse proportionality of the penetration length the magnetic
field implies the presence of a spontaneous surface magnetization weakly
dependent on $H$.

\section{Conclusion}

We have examined the paramagnetic contribution from surface zero-energy Andreev
bound states to the low-temperature penetration length of $d$-wave
superconductors in the Meissner state. The paramagnetic current is localized
within several coherence lengths near the surface and grows larger in the clean
limit when the temperature goes down. A broadening of the bound states chokes
their contribution and determines their actual role in shaping the penetration
length. We found that the upturn in the low temperature penetration depth lies
at $T_{m0}\sim \sqrt{\xi_0/\lambda_0}T_c$ in the clean limit where the
paramagnetic contribution from the bound states can be handled with
perturbation theory same as small low-temperature corrections to the
penetration depth from the screening current. The minimum broadening capable of
straightening out the upturn is $\gamma\approx T_{mo}$.

Furthermore, we examined the penetration depth when
the bound states must be kept track of beyond perturbation
theory. A divergence of $\lambda(T)$ was found at the
phase transition to a state with
spontaneous surface supercurrent. This transition occurs only with
smallish broadening, $\gamma< (\xi_0/\lambda_0)T_c$. In the clean
limit\cite{clean} and at low temperatures, there is a nonlinear regime
of the paramagnetic current already in magnetic fields substantially weaker
than the fields for the nonlinear
effects to show up in response of shielding supercurrents. The
broadening of the bound states modifies and weakens the nonlinear response.

Specifying an origin of the broadening as associated with nonmagnetic impurity
scattering, we obtained restrictions on the mean free path admitting the low
temperature anomalies. The conditions turn out to be sensitive to the strength
of the impurity potential and very different in the unitary and in the Born
limits. The Born impurities are shown to easily prevent the anomalies of the
penetration depth taking place at least well below $T_{m0}$. By contrast,
unitary scatterers with sufficiently small normal-state scattering rate
$\Gamma_u\ll T_c$ admit the transition to a state with spontaneous surface
supercurrent at $T_s\sim (\xi_0/\lambda_0)T_c$. In the latter case, however,
surface roughness very probably dominates the broadening and controls the bound
state contribution to the low-temperature penetration length.

\section*{Acknowledgments}

We thank M.~Fogelstr\"om for useful discussions.
This work was supported by the Academy of Finland, research Grant No. 4385.
Yu.S.B. acknowledges the financial support of Russian Foundation for Basic
Research under grant No.~99-02-17906.





\begin{references}
\bibitem{clean}
Throughout the article under the clean limit we mean the possibility to
disregard the broadening of the zero-energy bound states. The condition
$\xi_0\ll l$ is always assumed to be valid.
\end{references}
\end{document}